\documentclass[11pt,english]{article}
\usepackage[T1]{fontenc}
\usepackage[latin9]{inputenc}
\usepackage[a4paper]{geometry}
\usepackage{stmaryrd}
\usepackage{graphicx}
\usepackage{setspace}
\makeatletter
\newcommand{\lyxaddress}[1]{
\par {\raggedright #1
\vspace{1.4em}
\noindent\par}
}

\usepackage{siunitx}
\usepackage{cite}
\usepackage{hyperref}
\hypersetup{colorlinks,allcolors=blue}
\date{}

\makeatother

\usepackage{babel}
\begin{document}

\title{CMOS Platform for Atomic-Scale Device Fabrication}

\author{Tomáš Škere\v{n}$^{1}$, Nikola Pascher$^{1}$, Arnaud Garnier$^{2}$,
Patrick Reynaud$^{2}$,\\
Emmanuel Rolland$^{2}$, Aurélie Thuaire$^{2}$, Daniel Widmer$^{1}$,\\
Xavier Jehl$^{3}$, Andreas Fuhrer$^{1}$}
\maketitle

\lyxaddress{$^{1}$IBM Research \textendash{} Zurich, Säumerstrasse 4, 8803 Rüschlikon,
Switzerland\\
$^{2}$Commissariat à l\textquoteright Energie Atomique, Laboratoire
d\textquoteright Électronique des Technologies de l\textquoteright Information,
F38054 Grenoble, France and Université Grenoble-Alpes, F38000 Grenoble,
France\\
$^{3}$Commissariat à l\textquoteright Energie Atomique, Institut
Nanosciences et Cryogénie, F38054 Grenoble, France and Université
Grenoble-Alpes, F38000 Grenoble, France}
\begin{abstract}
Controlled atomic scale fabrication of functional devices is one of
the holy grails of nanotechnology. The most promising class of techniques
that enable deterministic nanodevice fabrication are based on scanning
probe patterning or surface assembly. However, this typically involves
a complex process flow, stringent requirements for an ultra high vacuum
environment, long fabrication times and, consequently, limited throughput
and device yield. Here, a device platform is developed that overcomes
these limitations by integrating scanning probe based dopant device
fabrication with a CMOS-compatible process flow.\textbf{ }Silicon
on insulator substrates are used featuring a reconstructed Si(001):H
surface that is protected by a capping chip and has pre-implanted
contacts ready for scanning tunneling microscope (STM) patterning.
Processing in ultra-high vacuum is thus reduced to only a few critical
steps which minimizes the complexity, time and effort required for
fabrication of the nanoscale dopant devices. Subsequent reintegration
of the samples into the CMOS process flow not only simplifies the
post-processing but also opens the door to successful application
of STM based dopant devices as a building block in more complex device
architectures. Full functionality of this approach is demonstrated
with magnetotransport measurements on degenerately doped STM patterned
Si:P nanowires up to room temperature.
\end{abstract}

\section{Introduction}

\textbf{}An early demonstration of the potential of scanning probe
microscopy (SPM) in the area of atomic scale fabrication was the famous
quantum coral \cite{Crommie1993}. Since then, a large number of distinct
probe-based fabrication methods have been developed based either on
direct atomic assembly or tip induced patterning \cite{Meyer2015}.
However, broader adoption of these methods is often impeded by their
inability to compete in speed and throughput with conventional lithography
techniques and to find a packaging approach that allows preservation
of the atomic scale devices outside an ultra high vacuum (UHV) environment. 

\begin{figure}
\begin{centering}
\includegraphics[width=16cm]{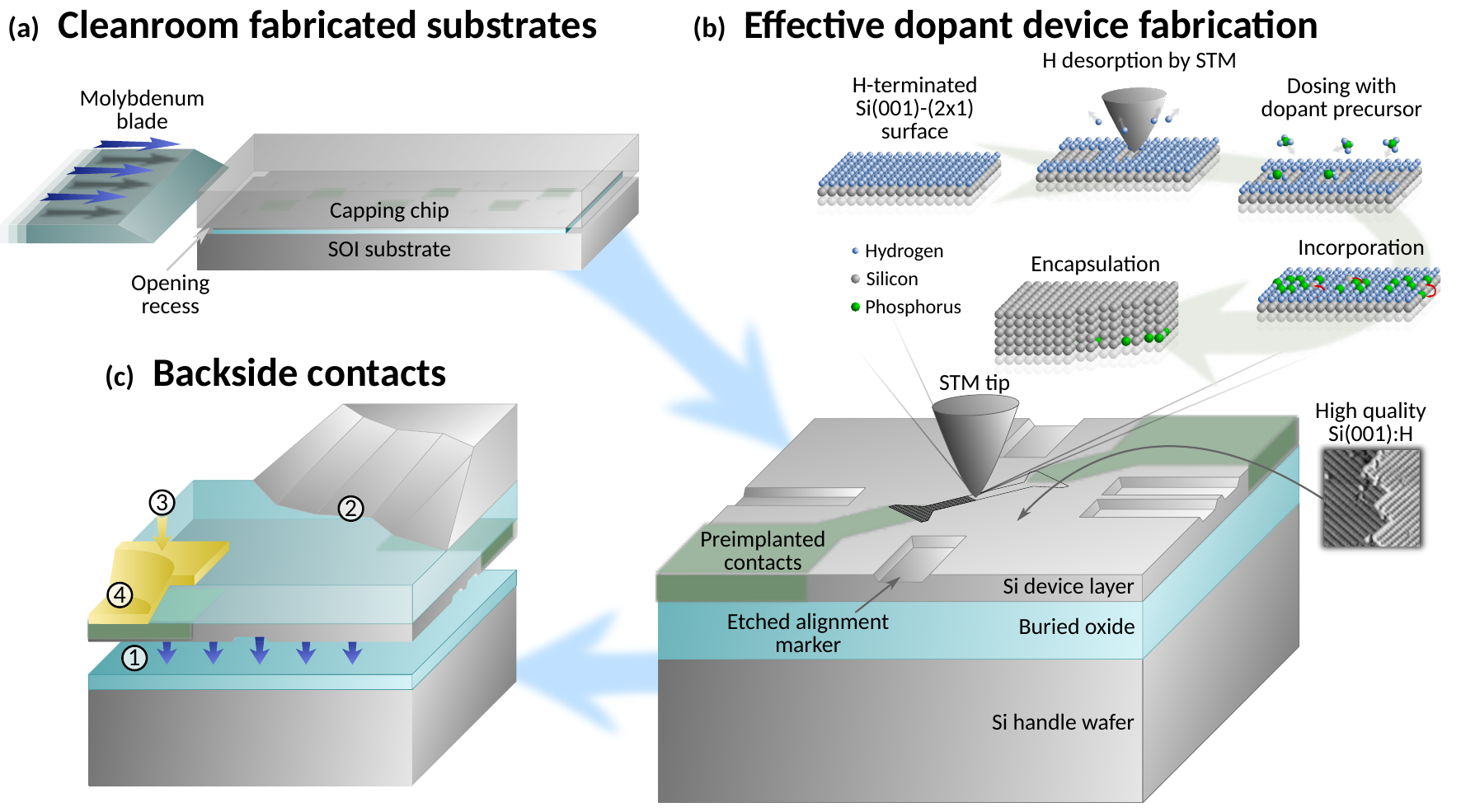}
\par\end{centering}
\caption{\label{fig:Introduction}SOI-based platform for atomic scale dopant
device fabrication. (a) Hydrogen passivated SOI substrates with pre-implanted
contacts and alignment markers protected from the ambient environment
by a similarly passivated capping chip that is hydrogen bonded to
the sample. After loading the sample into UHV, the capping chip is
removed with a molybdenum blade exposing a high quality Si(001):H
surface ready for STM patterning. (b) The STM tip is aligned to the
pre-implanted contacts with the help of etched alignment markers and
a desired pattern is written by locally desorbing the hydrogen from
the surface. Precursor gas ($\mathrm{PH_{3}}$) selectively sticks
to the dehydrogenated surface creating highly doped areas with defined
geometry. After the incorporation anneal and encapsulation of the
dopant device with a thin layer of Si, electrical contacts are fabricated.
(c) Chip is bonded to a carrier wafer (1), original handle wafer is
removed in a chemical mechanical process (2), and metal connections
(3) are fabricated through vias in the buried oxide layer (4). }
\end{figure}

One of the most promising SPM-based fabrication techniques is scanning
tunneling microscopy (STM) based dopant device fabrication \cite{Ruess2004}.
In this case, a hydrogen passivation layer on Si(001):H is locally
desorbed with an STM tip in a current induced process \cite{Shen1995,Becker1990}
to create reactive sites for selective attachment of dopant precursor
molecules (see Fig. \ref{fig:Introduction}b and Supplementary Information).
After thermal incorporation of the dopants into surface substitutional
sites, overgrowth with epitaxial silicon, and ex-situ contacting of
the dopant device layer, these atomically precise devices can be electrically
measured in the same way as conventional semiconductor devices. The
technique provides a way of defining degenerately doped metallic regions
in a semiconductor with atomic-scale resolution. It allows deterministic
placement of single dopant atoms \cite{Fuechsle2012,Schofield2003},
creation of wires \cite{Weber2012} and quantum dots \cite{Pascher2016,Fuhrer2009}
with atomic scale dimensions and enables exploration of the quantum
properties of these nanoscale systems. Furthermore, it may uniquely
provide the necessary atomic precision needed for donor based quantum
computing architectures \cite{OBrien2001,Oberbeck2002,Hile2015}.

However, as with other probe-based fabrication schemes, STM dopant
device fabrication is technically extremely challenging. It requires
the use of dedicated UHV equipment and the process flow for the device
fabrication is very time consuming, tedious and prone to failure.
As a result, only a handful of groups in the world have successfully
made devices with this technique \cite{Ruess2004,Bussmann2015,Pascher2016,Gramse2017},
and despite efforts directed at simplifying some parts of the process
\cite{Shen2004,Ward2017} its current application potential remains
somewhat limited.

To overcome these limitations we present a platform for STM based
dopant device fabrication that is fully integrated with a CMOS process
flow. It drastically reduces the fabrication time and complexity,
and adds additional functionalities to STM fabricated dopant devices.
The key features of this platform are outlined in Fig.~\ref{fig:Introduction}.
Silicon on insulator (SOI) substrates are prepared at wafer scale
(200\,mm) in a cleanroom environment using an original integration
based on standard CMOS processes (see Methods). Wafers are diced into
samples containing localization markers and pre-implanted contacts.
Each sample is protected with a sacrificial Si capping chip, which
can be easily removed inside the UHV system by inserting a molybdenum
blade in the recess between the substrate and the cap. Both chips
have reconstructed and hydrogen terminated Si(001):H surfaces with
low defect densities directly suitable for STM patterning without
the need for high-temperature in-situ surface preparation. Large implanted
contact pads serve as easy-to-reach contact terminals for post-processing
while shallow implant extensions come as close as 670\,nm from each
other and are easily accessible in the STM patterning process. After
UHV fabrication and Si overgrowth, the devices can either be contacted
from the top using optical or e-beam lithography or the chips can
be re-integrated into the CMOS workflow using a chip-to-wafer bonding
process (Fig. \ref{fig:Introduction}c). In the latter case, many
devices can be simultaneously contacted from the backside by removal
of the silicon substrate and contact can be made to the device through
vias in the buried oxide (BOX) of the sample chips. Moreover, the
shallow thickness (200\,nm) of the SOI device layer eliminates current
leakage through the substrate and allows electrical operation of STM-defined
dopant devices even at room temperature. 

\begin{figure}
\begin{centering}
\includegraphics[width=8.4cm]{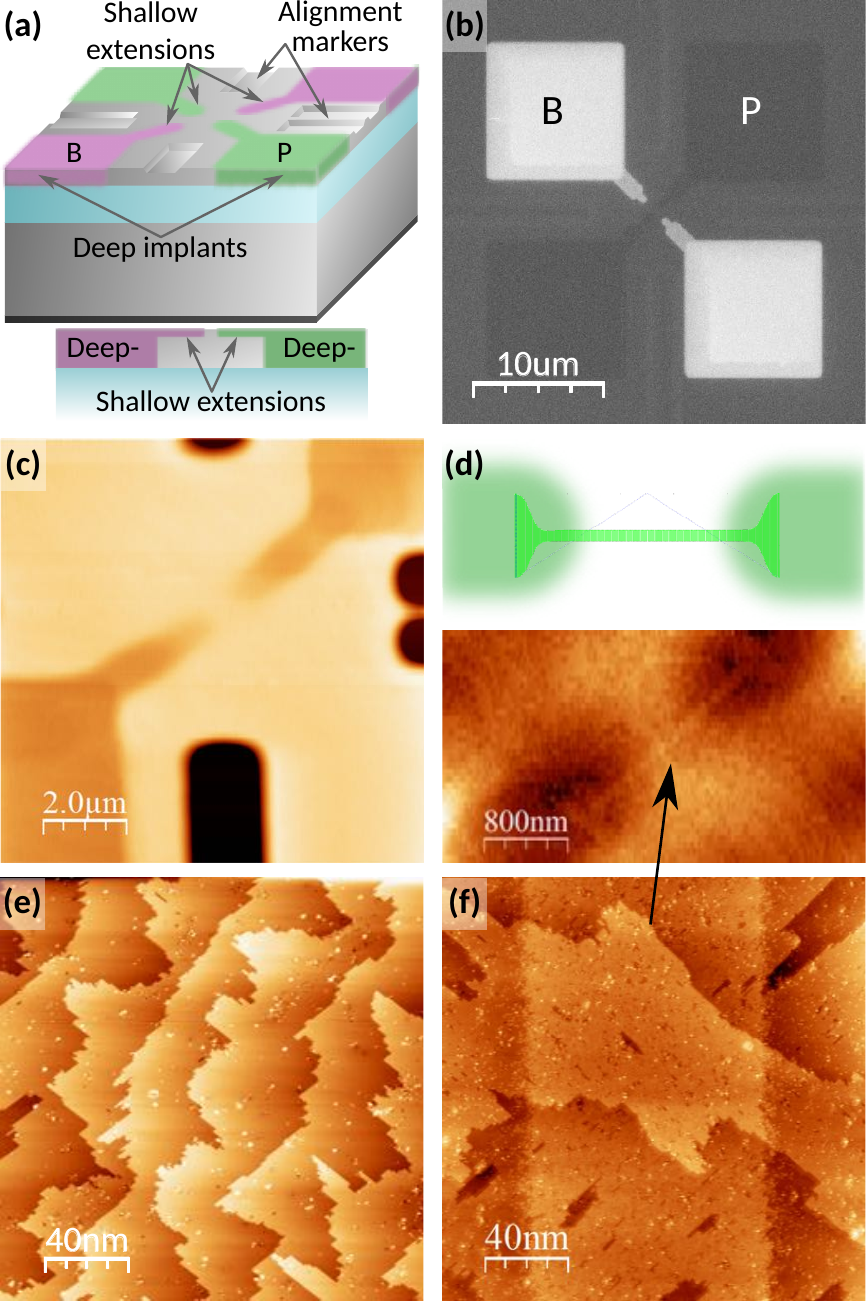}
\par\end{centering}
\caption{\label{fig:Process-flow-of-STM_litho}STM-based dopant device fabrication
on an SOI sample with pre-implanted contacts. (a) Schematic of the
central device region of the SOI samples with both boron (B) and phosphorus
(P) implants. (b) Scanning electron microscope image of the device
region with both types of implanted contacts each giving a different
contrast. (c) STM image of two n-type contacts (top right to bottom
left) and the three etched alignment markers (dark areas). (d) High
resolution STM image of the device area showing the typical surface
quality of the SOI substrates, with few residual defects that result
from the de-bonding process. (e) STM image of the device area with
a desorbed 30\,nm wide nanowire connecting the two contacts and (f)
schematic drawing of the nanowire with dimensions. }
\end{figure}

\section{Methods}

\subsection*{SOI substrate fabrication}

The pre-fabricated SOI substrates were made in a cleanroom at CEA
LETI. A detailed process flow is provided in the Supplementary Information.
We start with a 200\,mm SOI wafer with a 200\,nm thick device layer
(B doped, resistivity $\sim$10\,$\Omega$\,cm) and 400\,nm thick
buried oxide (BOX). In the first step we etch the alignment marks.
They are 50\,nm deep and sidewalls are at an angle of \ang{55} which
provides good marker visibility but does not represent a problem for
scanning with the STM tip. 

Contact implantation is performed in a two step process: the deep
implants (fluence of \SI{3e15}{cm^{-2}} and ion energy of 20\,keV
and 80\,keV for P and B, respectively) extend all the way through
the device layer and allow contacting from the backside as required
for CMOS reintegration \cite{Thuaire2016}. In contrast, the shallow
implants (fluence of \SI{3e15}{cm^{-2}} and ion energy of 5\,keV
and 12\,keV for P and B, respectively) only create a 20\,nm deep
doped layer (before reconstruction) and limit damage to the surface
due to the implantation process. 

In order to achieve sufficient surface quality, retain marker geometry
and limit dopant diffusion near the contact pads we use a reduced
pressure chemical vapor deposition (CVD) process \cite{Hartmann2002}
for surface reconstruction and hydrogen termination of the SOI substrates.
The surface reconstruction step is \textit{\emph{performed}} at a
temperature of \SI{950}{\celsius} which is sufficiently low to prevent
substantial dopant diffusion. Finally, we protect the reconstructed
surface by hydrophobic bonding with a sacrificial capping wafer \cite{Rauer2013}.
The sandwich is then diced and the chips can be introduced into UHV
for dopant device fabrication. Once capped, the samples can be kept
in ambient conditions for many months without any noticeable degradation
of the surface quality.

\subsection*{Dopant device fabrication with STM lithography}

In order to handle the SOI samples in the UHV STM setup we designed
a specialized sample holder that clamps the substrate from the sides
allowing for easy cap removal in UHV. After introducing the sample
into the system, it is first degassed for 1\,h at \SI{300}{\celsius}
in order to remove adsorbed chemical species. The capping chip is
removed by pressing a molybdenum blade in a recess between the sample
and the cap (see Fig. S2 in the Supplementary Information). Finally,
the sample is transferred to the STM stage and lithography is performed. 

For this we use a commercial variable temperature STM from Omicron
with custom control software and hardware. Alignment markers are used
to position the tip close to the central device area which contains
the contact implants (Fig. S3-5 in the Supplementary Information). 

For STM imaging of the samples we typically use a sample bias of -2.5\,V
and tunneling current setpoint of 100\,pA. In order to desorb hydrogen
from the surface, we switch to writing parameters and scan over the
desired areas with the feedback switched on. For patterning we apply
a positive sample bias between +4.5\,V and +7\,V and tunneling current
of around 4\,nA. For slow writing of fine features we use tip velocity
down to \SI{10}{\nm\per\s}, separation of lines in the desorption
pattern of $\sim$0.5\,nm and voltage from the lower part of the
range (+4.5\,V to $\sim$+5.5\,V). For rapid, coarse patterning
of large areas we use tip velocity up to \SI{100}{\nm\per\s}, line
separation of $\sim$5\,nm and voltage up to +7\,V. We use Python
scripts to generate the desired patterns for desorption and to programmatically
control the STM tip movement along the pattern.

After STM desorption of the pattern is complete, we typically image
the desorbed area (or a part of it) in order to confirm desorption
quality and check for desorption errors. Then we proceed with a doping
procedure (similar to the one described in \cite{Pascher2016}). The
sample is exposed to a 10\,L dose of phosphine gas while still in
the STM stage. Afterwards, the sample is transferred to another chamber
(of the same UHV system), placed on a heated manipulator and annealed
to \SI{370}{\celsius} for 1\,min. After cooling the sample to a
temperature of about \SI{270}{\celsius}, a 20\,nm thick layer of
intrinsic Si is grown on top of the sample using a silicon sublimation
source from MBE-Komponenten GmbH at growth rate of about 1\,nm/min.

\subsection*{Sample post-processing}

Compared to the conventional dopant device fabrication on bulk Si
substrates, the contact fabrication procedure is facilitated by the
presence of the large (\SI{10x10}{\micro\m}) pre-implanted contact
pads. There are two possibilities for the post-processing: I) direct
e-beam lithography (or, alternatively, optical lithography) contacting
on a single chip level or II) reintegration of the samples in the
wafer scale CMOS process flow (see Supplementary Information). 

I) For the \textbf{e-beam process}, metal contacts are fabricated
directly on the upper side of the samples. After overgrowth, the implanted
areas are buried under a 20\,nm thick layer of intrinsic Si. We first
use reactive ion etching (RIE) to etch a grid of $1\times1$\,$\mathrm{\mu m^{2}}$
large and 50\,nm deep holes into the implant pads and subsequently
deposit 100\,nm thick Al contacts in a standard lift-off process.
The contacts are then annealed to \SI{350}{\celsius} for 15\,min
in a reducing atmosphere (5\% H, 95\% Ar) at 200\,mbar. In order
to minimize leakage between the handle wafer (used as a gate in this
case) and the device layer, we etch a frame around each device position
isolating the active area from the rest of the device layer (particularly
from the edges of the sample which often cause leakage). Finally,
the sample is glued into a chip carrier using silver epoxy (Epotec
H20e) and wire-bonded. In order to apply a gate voltage to the handle
wafer, the back side of the sample is scratched with a diamond scribe
prior to gluing in order to penetrate the oxide. 

II) For \textbf{CMOS reintegration}, the chips are first bonded to
a new handle wafer (bulk Si wafer with 200\,nm oxide layer on top)
with the device layer (where STM device is located) facing towards
the handle wafer. The original SOI substrate is then completely removed
using a chemical-mechanical process, exposing the backside of the
original BOX layer. In the next step we etch vias through the BOX
exposing the bottom side of the deep contact implants in the device
layer. Finally, we deposit aluminum silicon alloy contacts which connect
the implants through the vias to wire-bonding pads. A custom lithography
step was performed on the \textit{30\,nm NW} sample to additionally
define a metallic top gate. This could have been done in the reintegration
step but was not foreseen in the mask-set that was used. Finally,
the sample is also glued in a chip carrier and wire-bonded. 

\subsection*{Electrical transport measurements }

Transport measurements on the samples were performed using standard
lock-in techniques (Signal recovery - 7265) in a four-point configuration.
Low temperature measurements were carried out in a custom made variable
temperature insert (VTI) cryostat allowing stable sample temperatures
in a range from 2\,K up to 300\,K. 

For the e-beam processed samples the handle wafer was used as a gate.
Due to its non-degenerate doping it becomes insulating at temperatures
below $\sim$30\,K. To circumvent this problem, we applied a short
(200\,ms) light pulse after every change of the gate voltage using
a red LED (0402 surface mount LED, 632\,nm peak wavelength) that
was mounted in the chip carrier. This creates enough charge carriers
to equalize the potential in the handle wafer and define a well defined
voltage at the BOX interface. This was not necessary for the reintegrated
samples which were equipped with a metallic gate. 

\section{Results and discussion}

In the STM stage the pre-implanted contacts are localized using an
optical microscope and large area scans that typically take less than
30 minutes (see Supplementary Information). A schematic of the device
position with pre-implanted contacts is shown in Fig.~\ref{fig:Process-flow-of-STM_litho}a,
whereas Fig.~\ref{fig:Process-flow-of-STM_litho}b shows a scanning
electron microscopy (SEM) image of the device position. Fig.~\ref{fig:Process-flow-of-STM_litho}c
shows a constant current topographic STM image taken just before defining
a nanowire between two n-type contacts. The implants are clearly visible
thanks to a height difference of a few monolayers (5-10) caused by
different oxidation and etch rates for heavily doped Si in the surface
reconstruction process after the ion implantation step. The typical
quality of the Si(001):H sample surface in the central area is indicated
by the STM image in Fig.~\ref{fig:Process-flow-of-STM_litho}d.

Figure~\ref{fig:Process-flow-of-STM_litho}e shows the center of
the same device position with a 30\,nm nanowire defined by hydrogen
desorption (see Methods) with the STM tip. A dimensional representation
of the nanowire pattern which was used is shown in Fig.~\ref{fig:Process-flow-of-STM_litho}f.
The duration of the writing process in this case was about 15\,minutes.
After the writing step the sample was dosed with phosphine, annealed
at \SI{370}{\celsius} for one minute to incorporate and electrically
activate the dopants, overgrown with 20\,nm of intrinsic Si and then
removed from UHV (see Methods). This specific nanowire device was
subsequently reintegrated into the 200\,mm CMOS process flow for
backside contact fabrication. In the following this device is referred
to as \emph{30\,nm NW}. 

In a similar way we fabricated a 120\,nm wide nanowire with the same
length of 670\,nm. This second sample was post-processed using e-beam
lithography and by contacting the large implant pads from the front.
This sample is referred to as \emph{120\,nm NW}. A third sample containing
no STM defined structures was post-processed together with the \textit{120\,nm
NW} sample. Two device positions with an implant separation of 670\,nm
and 1170\,nm were contacted and are referred to as \emph{Blank 1}
and \emph{Blank 2}, respectively. 

\textbf{}An important advantage of the SOI substrate is that the
charge carrier density in the device layer can be tuned by application
of a voltage to the silicon substrate. For samples contacted from
the top by e-beam lithography we find that the presence of trapped
charges at the BOX interface shifts the Fermi level in a way that
leakage occurs between contacts in the device layer. By application
of a negative voltage to the substrate this can be fully suppressed
and the charge compensated. 

\begin{figure}
\begin{centering}
\includegraphics[width=16cm]{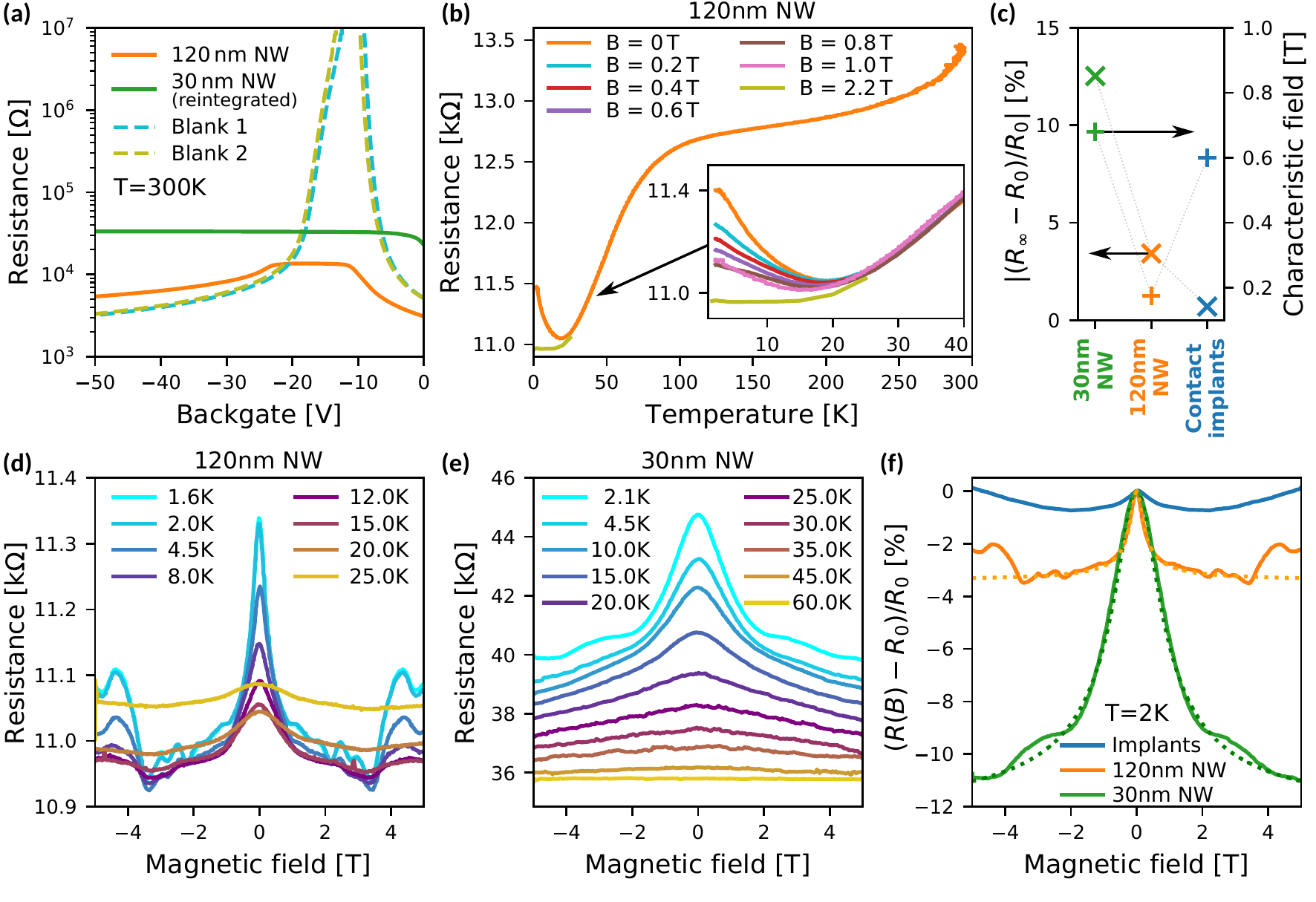}
\par\end{centering}
\caption{\label{fig:Electrical_measurements}Electrical measurements of dopant
devices fabricated on SOI substrates. (a) Room temperature resistance
measurements of the \textit{120\,nm NW} and \textit{30\,nm NW} and
two corresponding \textit{Blank} positions as a function of gate voltage
(b) Resistance of the \textit{120\,nm NW} from (a) as a function
of temperature from 2\,K up to room temperature. Inset shows the
temperature dependence measured in magnetic fields of different values.
(c) Comparison of heights ($\left|\left(R_{\infty}-R_{0}\right)/R_{0}\right|$)
and widths (characteristic field) of the weak localization peaks from
(e). (d) Magnetoresistance of the \textit{120\,nm NW} and (e) \textit{30\,nm
NW} measured at different temperatures. (f) Comparison of magnetoresistances
of the \textit{120\,nm NW }and reintegrated \textit{30\,nm NW} with
the magnetoresistance of implanted contacts. }
\end{figure}

In Fig.~\textbf{\ref{fig:Electrical_measurements}}a we present room
temperature measurements of the electrical transport through the nanowires
and \textit{Blank} positions as a function of the applied gate voltage.
In the case of the \textit{120\,nm NW} and \textit{Blank }positions
the gate was realized by applying a voltage to the handle wafer, while
for the reintegrated \emph{30\,nm NW} a metallic gate was lithographically
defined in the post-processing step (see Supplementary Information).
For the three e-beam contacted samples, the trapped charges in the
BOX shift the layer into inversion. Application of a negative bias
of -10\,V leads to a depletion of the device layer and fully insulating
behavior even at room temperature. This uniquely enables the measuring
of electrical transport through STM defined dopant devices at room
temperature. The situation is similar for the \emph{30\,nm NW}; however,
after removal of the handle wafer in the reintegration process the
Fermi level offset is not as pronounced as for the other samples and
even at zero gate bias substrate leakage is small. 

Figure~\ref{fig:Electrical_measurements}b shows the resistance
of the \textit{120\,nm NW} as a function of temperature. The resistance
drops with decreasing temperature as expected for a metallic system
and at temperatures below 25\,K it increases again due to the onset
of weak localization (WL) \cite{Beenakker1991}. This regime is governed
by enhanced backscattering from constructive interference of time-reversed
paths with a length that is shorter than the phase coherence length.
Resistivity is therefore sensitive to time-reversal symmetry breaking
by a magnetic field and the increase in resistance is suppressed for
high enough fields \cite{AlTshuler1981}. Figures~\ref{fig:Electrical_measurements}d
and e show in more detail the magnetoresistance of the two nanowires
at different temperatures. Aside from the WL peak around $B=\mathrm{0\thinspace T}$,
the resistance of the nanowires (particularly the \textit{120\,nm
NW}) is also affected by universal conductance fluctuations \cite{Beenakker1991}.
The relative magnitude of the WL effect in the two nanowires at 2\,K
is compared in Fig.~\ref{fig:Electrical_measurements}f. The dotted
lines are fits of the peaks to the 1D WL theory (see Supplementary
Information). The resistivity of the \emph{30\,nm NW} and \textit{120\,nm
NW} is $1.74\thinspace\mathrm{k\Omega/\boxempty}$ and $1.96\thinspace\mathrm{k\Omega/\boxempty}$
and the phase coherence length based on fits to WL theory is 56\,nm
and 54\,nm, respectively. This is comparable to previously reported
values \cite{Ruess2007}. The blue curve shows the situation for a
reference sample with no gap between the implanted contacts where
the dopants form a 3D channel with a crosssection of about 700$\times$100\,nm.
Figure \ref{fig:Electrical_measurements}\textbf{c} compares the magnitude
and characteristic field of the WL effect for the three different
cases. As expected, the magnitude reduces as the width and thickness
of the dopant layer increases. The characteristic phase breaking field
is, however, larger again for the implant case, since only the projection
of the 3D scattering paths onto a plane perpendicular to the field
is relevant for phase breaking.\textbf{} The electrical transport
measurements of the dopant nanowires show consistent behavior similar
to previous reports and confirm that the suggested alternative strategy
for sample pre- and post-processing is compatible with the sensitive
STM-based dopant device fabrication process. 

\textbf{}

\section{Conclusions}

In summary, we have developed a platform which integrates STM-based
nanoscale dopant device fabrication in the standard CMOS process flow.
Performing STM lithography on pre-implanted and hydrogen terminated
SOI substrates greatly reduces the fabrication time and complexity
and it facilitates sample post-processing. The thin SOI device layer
suppresses substrate current leakage and, with the assistance of a
gate, allows room temperature operation of the dopant devices. Last
but not least, the demonstrated CMOS compatibility opens the possibility
to integrate atomic scale dopant devices with additional on-chip circuitry
and may in the future be extended to other STM fabricated devices
such as the ones resulting from molecular assembly. 

\section*{Acknowledgements}

The authors gratefully acknowledge financial support from EU-FET grants
SiAM 610637, PAMS 610446 and from the Swiss NCCR QSIT. 


\end{document}


\title{Supplementary Information for:\\
CMOS Platform for Atomic-Scale Device Fabrication}

\author{Tomáš Škere\v{n}$^{1}$, Nikola Pascher$^{1}$, Arnaud Garnier$^{2}$,
Patrick Reynaud$^{2}$,\\
Emmanuel Rolland$^{2}$ Aurélie Thuaire$^{2}$, Daniel Widmer$^{1}$,\\
Xavier Jehl$^{3}$, Andreas Fuhrer$^{1}$}
\maketitle

\lyxaddress{$^{1}$IBM Research \textendash{} Zurich, Säumerstrasse 4, 8803 Rüschlikon,
Switzerland\\
$^{2}$Commissariat à l\textquoteright Energie Atomique, Laboratoire
d\textquoteright Électronique des Technologies de l\textquoteright Information,
F38054 Grenoble, France and Université Grenoble-Alpes, F38000 Grenoble,
France\\
$^{3}$Commissariat à l\textquoteright Energie Atomique, Institut
Nanosciences et Cryogénie, F38054 Grenoble, France and Université
Grenoble-Alpes, F38000 Grenoble, France}

\section{\label{sec:Conventional-STM-lithography}Conventional STM lithography
procedure}

In this section we discuss the usual procedure for dopant device fabrication
on bulk Si substrates, we highlight the most time-consuming and throughput-limiting
steps and compare it to the newly developed method on silicon-on-insulator
(SOI) samples that was presented in the main text of the paper. 

\subsection{Sample preparation}

On a wafer scale, optical localization markers are defined using e-beam
lithography and reactive ion etching. The wafer is then diced into
$3\times9\;$mm sized sample chips. After a standard ex-situ cleaning
procedure (piranha clean for 10\,min at a temperature of \SI{80}{\celsius},
removal of the oxide using a 10\,s dip in buffered hydrofluoric acid
solution 7:1 and a final RCA-2 clean) the Si sample is introduced
into the ultra-high vacuum (UHV) system, degassed at \SI{500}{\celsius}
for about 10\,h, flashed to \SI{1150}{\celsius} three times before
hydrogen passivation in an atomic hydrogen beam (MBE Komponenten HABS
II) at $325^{\circ}\mathrm{C}$. The sample is then transferred to
the scanning tunneling microscope (STM) stage for dopant device fabrication.
Total preparation time without the fabrication and cleaning of the
sample chips takes approximately 12 hours. 

In the case of the SOI substrates, very little in-situ preparation
is required. After loading the sample into UHV, the capping chip is
removed, a short 1\,h degas at \SI{300}{\celsius} is performed and
the total time between loading the sample and STM lithography is less
than 2 hours. 

\subsection{STM lithography}

\begin{table}[b]
\begin{centering}
\begin{tabular}{|c|c|c|c|c|}
\hline 
Mode & Voltage {[}V{]} & Current {[}pA{]} & Speed {[}nm.s$^{-1}${]} & Line width {[}nm{]}\tabularnewline
\hline 
\hline 
Imaging & $-$(1.9-2.5) & 50-500 & 300 & -\tabularnewline
\hline 
Writing (slow, high resolution) & $+$4-5.2 & 2000-4000 & 10-50 & 0.5-2\tabularnewline
\hline 
Writing (fast, low resolution)  & $+$5-7 & 2000-4000 & 50-100 & 5-10\tabularnewline
\hline 
\end{tabular}
\par\end{centering}
\caption{\label{tab:Typical-STM_param}Typical imaging and writing STM parameters
(bias measured on the sample with respect to the tip)}
\end{table}

As soon as the sample is in the STM stage, the desired areas are desorbed
with the STM tip according to a predefined meander pattern. The typical
STM parameters we use for scanning and STM lithography are summarized
in Table \ref{tab:Typical-STM_param}. STM based hydrogen desorption
of the active area of the nanodevice, which is typically less than
a few hundred nanometers in size, only takes a few minutes or up to
an hour for larger and more complex device designs. In the second
step, a large contact pad has to be patterned for each electrical
terminal, that can later be localized in scanning electron microscopy
(SEM) and contacted with the e-beam lithography. Typical size of these
pads is \SI{1x2}{\micro\m} and the time required for the STM desorption
is around 1-2 hours per pad. Including imaging, all required realignment
steps and drift stabilization, writing of a typical dopant device
therefore takes 6-8 hours. 

Using the new SOI platform requires an additional initial step - precise
localization of the device position and implanted contacts. This procedure
is described in detail later in section \ref{sec:Alignment-and-STM}
and the typical duration is less than 30\,min. Once the implanted
contacts are localized, only the active device area is desorbed and
no contact pads need to be defined. This shortens the entire STM writing
procedure to about an hour and allows multiple devices to be written
on one chip in a single run. 

\subsection{Dosing, activation anneal and overgrowth}

After the STM lithographic step, the sample is dosed with the dopant
precursor (typically phosphine at $p=$\SI{1.0e-7}{mbar} for 1\,min),
undergoes an activation anneal at \SI{350}{\celsius} for one minute
and then it is overgrown with a 20\,nm intrinsic silicon layer. The
substrate temperature during overgrowth is held at around \SI{270}{\celsius}
with the possibility to start with a so-called locking layer - to
grow the first few monolayers at a lower temperature in order to prevent
dopant segregation \cite{Keizer2015}. The typical growth rate is
on the order of 1\,nm/min. After overgrowth, the dopant device is
encapsulated in the Si matrix and can be exposed to the ambient environment. 

This procedure is essentially identical for the SOI samples and it
takes less than an hour including temperature ramps and sample transfers
between chambers. 

\subsection{\label{subsec:Sample-post-processing}Sample post-processing}

In order to contact the device we use e-beam lithography with several
steps. In the first step, the sample is imaged with SEM to roughly
localize the device. In the second step, platinum markers are defined
in the vicinity of the device, and it is imaged together with the
platinum markers in order to have an exact reference for its location.
Then we use reactive ion etching to create an array of $\sim$30\,nm
deep holes with a diameter of 60\,nm over the contact pads to penetrate
the intrinsic Si overgrowth layer. In the final step, we deposit a
90\,nm thick aluminum layer to electrically contact the device terminals.
The post-processing usually takes at least 3 days, possibly more,
depending on logistics and tool availability. Finally, the chip is
glued in a chip carrier and wire-bonded. 

Post-processing is significantly different in the case of SOI substrates
and is described in detail in section \ref{sec:Full-process-flow}
below. The samples can be contacted either by e-beam lithography or
can be reintegrated into the CMOS process flow. Both options have
advantages over the conventional process. In the case of e-beam processing,
no device imaging is necessary and the via etch and metal deposition
steps are substantially easier due to the large size of the implants
and their well defined position. This process could be conveniently
performed by optical lithography as well. The CMOS reintegration process
involves additional processing steps. On the other hand, it allows
routine parallel processing of tens of chips at the same time and,
most importantly, opens up a possibility to integrate additional circuitry
and functionality to the device chips. In the device run from which
data is presented in the main text, reintegration took place with
many device chips but only two of them had an actual STM device on
them. 

\section{\label{sec:Full-process-flow}Full process flow for dopant device
fabrication on SOI substrates }

\begin{figure}
\begin{centering}
\includegraphics[width=16cm]{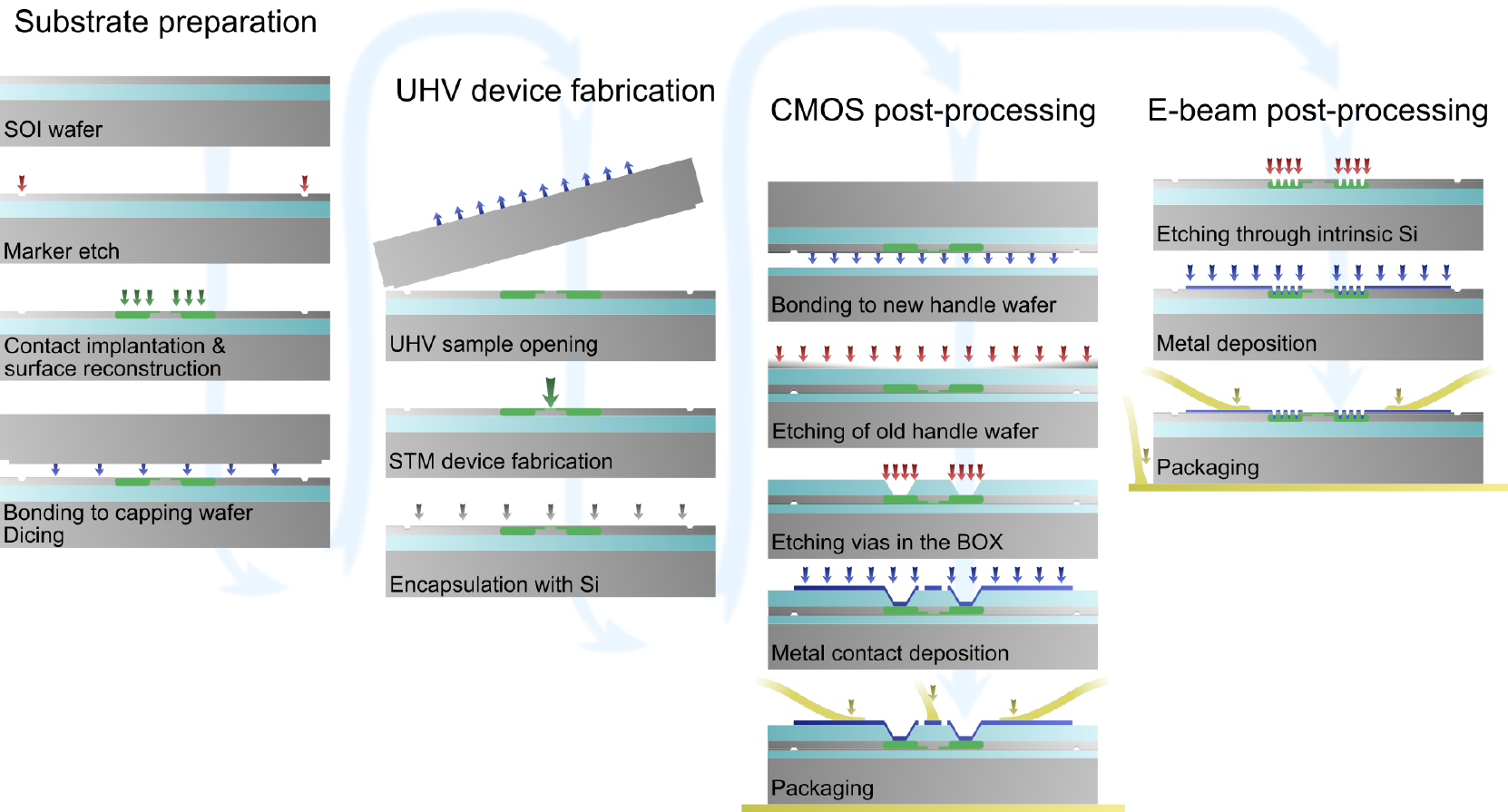}
\par\end{centering}
\caption{\label{fig:Full_process_flow}Outline of the process flow for dopant
device fabrication on SOI substrates. After substrate preparation
and capping, the sample is inserted in the UHV environment where the
capping chip is removed and the STM lithographic step is performed.
After encapsulation with 20\,nm of Si the sample is taken out of
the UHV environment and post-processed. Two distinct pathways are
possible \textendash{} reintegration of the samples in the CMOS process
by chip-to-wafer bonding and chip level postprocessing by e-beam (or
optical) lithography. }
\end{figure}

Figure \ref{fig:Full_process_flow} shows the full process flow of
the dopant device fabrication on SOI substrates including substrate
preparation and two alternative methods for postprocessing. 

The substrate preparation is performed on a wafer scale using Deep
UV lithography. The detailed structure of the markers and implants
is described in section \ref{sec:Layout-and-characterization}. After
the surface reconstruction step, the wafer is capped with another
wafer using hydrophobic bonding before dicing into sample chips. This
requires a fine-tuning of the bonding force to a such as to withstand
dicing but still make removal of the cap in UHV straightforward. With
the capping chip the sample surface is protected and samples can be
kept in ambient conditions basically indefinitely (sample surface
quality does not show any deterioration after about a one year period).
Once inserted into the vacuum, the capping chip is removed and STM
lithography is performed. 

The two distinct post-processing schemes are outlined in the third
and fourth column of Fig. \ref{fig:Full_process_flow}. 

In the case of CMOS post-processing, the samples are first chemically
oxidized and then bonded face-down to a new handle wafer using a hydrophilic
(oxide to oxide) bonding process (the handle wafer has a 200\,nm
layer of thermal oxide). In the next step, the original handle wafer
is etched away by a chemical-mechanical polishing (CMP) process, vias
penetrating the buried oxide (BOX) are etched, and metal (aluminum
silicon alloy) connections to the back side of the deep contact implants
are deposited. Finally, the sample is mounted in a chip carrier and
bonded. 

Alternatively, the sample containing the dopant devices can be processed
by standard e-beam lithography, which involves etching vias through
the Si overgrowth layer and depositing metal connections. Compared
to the conventional process described in section \ref{subsec:Sample-post-processing},
there are less lithographic steps involved and their accuracy is much
less critical. 

\section{\label{sec:Layout-and-characterization}Layout and characterization
of the SOI substrates}

\begin{figure}
\begin{centering}
\includegraphics[width=8cm]{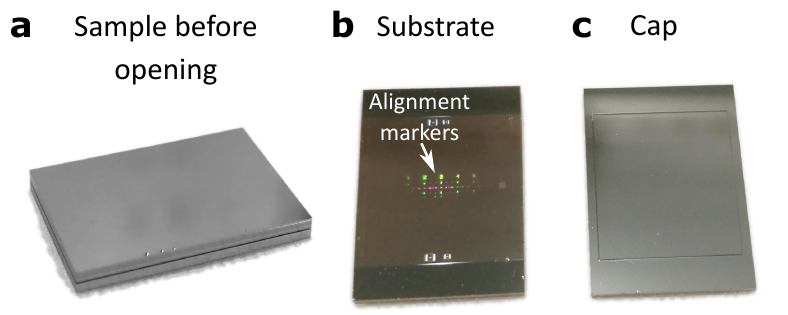}
\par\end{centering}
\caption{\label{fig:SOI_chips_photo}Photographs of the SOI substrate. \textbf{a,}
Sandwich structure before opening. \textbf{b,} The\textbf{ }substrate
with alignment marks visible and \textbf{c,} the capping chip. }
\end{figure}

\begin{figure}
\begin{centering}
\includegraphics[width=16cm]{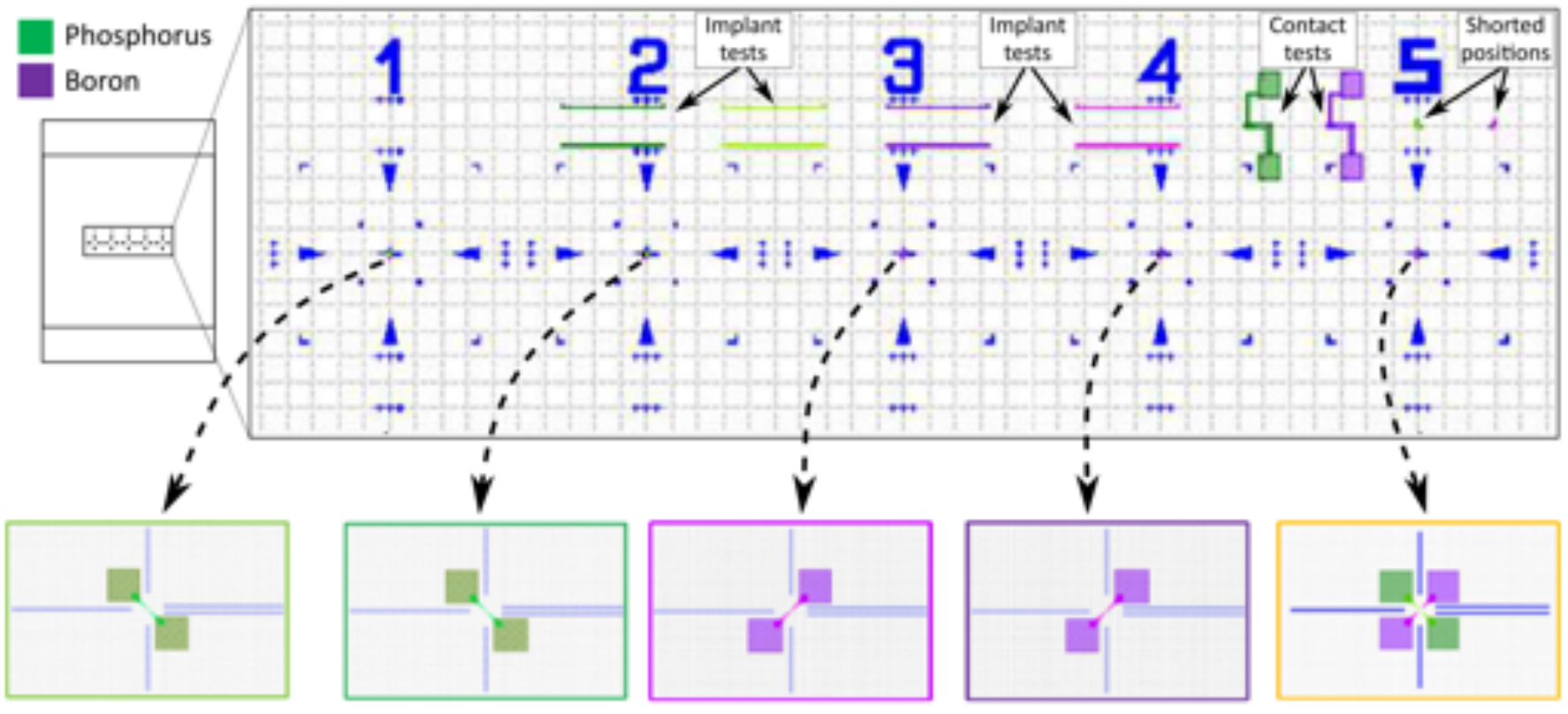}
\par\end{centering}
\caption{\label{fig:Localization_marker_layout}Detailed structure of the alignment
markers and implants on an SOI substrate including 5 device positions
and a number of test structures in the upper part. }
\end{figure}

Figure \ref{fig:SOI_chips_photo} shows the SOI substrate and the
capping chip before and after opening, with optical localization markers
visible. Figure \ref{fig:Localization_marker_layout} shows the layout
of the localization markers and the implants. Each chip contains five
device positions, two with a single pair of P implants, two with a
single pair of B implants and one compound position with a pair of
implants for both dopant types. Additionally, just below the position
numbers there is a row of various testing structures, including linear
implants and contact tests. Just below the number 5 there is a testing
structure which is identical to P doped device positions 1 and 2 but
with an implanted stripe connecting the two contacts. This testing
structure was also used to evaluate the magnetoresistance of the P
implants. 

\begin{figure}
\begin{centering}
\includegraphics[width=15cm]{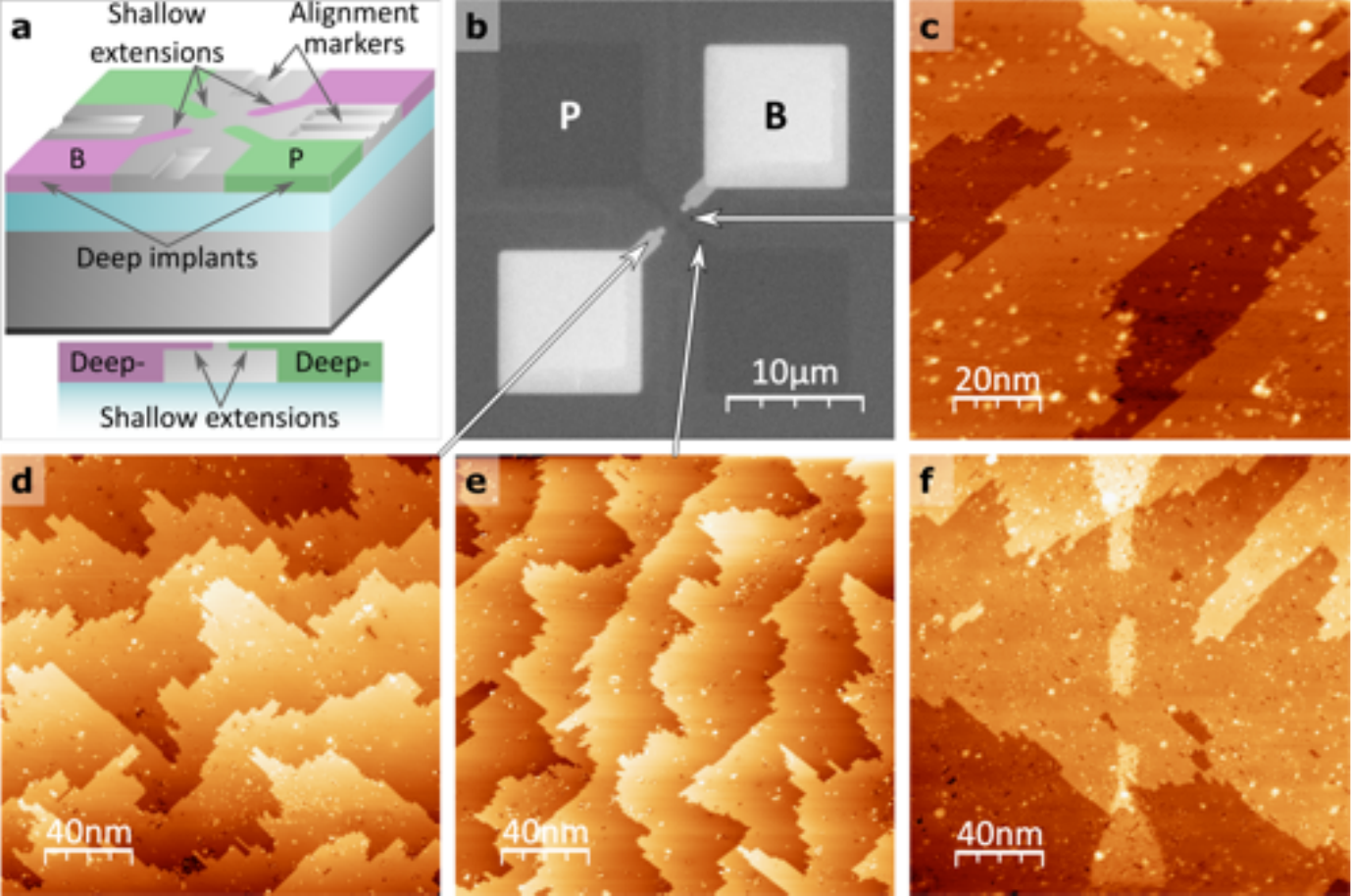}
\par\end{centering}
\caption{\label{fig:STM-surface}\textbf{a,} Structure and \textbf{b,} SEM
image of the implanted contact pads acquired using the in-lens detector
at $1.2~\mathrm{keV}$ primary beam energy. STM images of the sample
surface on the \textbf{d,} boron and \textbf{e,} phosphorus contact
pad extensions that were implanted at reduced ion energy. \textbf{c,}
STM image of the sample surface in the device area and \textbf{f,}
an example of STM patterning on the SOI substrate. }
\end{figure}

Figure~\ref{fig:STM-surface}\textbf{a} schematically depicts the
structure of the pre-implanted contacts and the etched localization
markers (the compound device position number 5). An actual SEM image
is shown in Fig.~\ref{fig:STM-surface}\textbf{b} with bright B-implanted
contacts and dark P-implanted contacts. The shallow implant extensions
run towards the center of the device position. The distance between
the opposite implants is 670\,nm (position 1, 3 and P implants in
position 5) or 1170\,nm (position 2, 4 and B implants in position
5).   

Figures~\ref{fig:STM-surface}\textbf{c}, \textbf{d} and \textbf{e}
show typical STM micrographs of the SOI sample surface. For the STM
dopant device fabrication it is critical that both the virgin silicon
surface in the center of the device area as well as the implanted
areas have a surface quality suitable for patterning. High implantation
energies and doses were found to be detrimental for the surface, which
is why we use a two-stage implantation process with high-energy deep
implants for the contact pads and low-energy shallow implant extensions
that reach closer to the the active device area (see Fig.~\ref{fig:STM-surface}\textbf{a}).
The deep implants (fluence of \SI{3e15}{cm^{-2}} and ion energy of
20\,keV and 80\,keV for P and B, respectively) extend all the way
through the device layer and enable contacting from the backside as
required for CMOS reintegration \cite{Thuaire2016}. In contrast,
the shallow implants (fluence of \SI{3e15}{cm^{-2}} and ion energy
of 5\,keV and 12\,keV for P and B, respectively) only create a 20\,nm
deep doped layer and preserve the atomic scale quality of the reconstructed
surface. The STM images in Fig.~\ref{fig:STM-surface}\textbf{c},
\textbf{d} and \textbf{e} were taken on the bare silicon and shallow
P and B implants, respectively. We find the surface quality to be
comparable in all three images and conclude that the shallow ion implantation
does not impact surface quality on the level required for STM patterning.
On the other hand, we found that surface reconstruction and hydrogen
termination of the capping chip is critical for preserving atomic
flatness of the sample surface. It is worth mentioning that even though
we tried to tightly fix all the process parameters and waiting times,
i.e., the time between surface reconstruction and bonding of the capping
wafer, we still found a noticeable variability of the surface quality
between different runs. This seems to indicate that extreme care during
wafer handling is critical. The quality of multiple samples from a
single wafer was, however, nearly identical.

The ultimate test for surface quality is, in our case, the ability
to desorb the hydrogen resist with the STM. Figure~\ref{fig:STM-surface}\textbf{f}
shows an example of a desorbed pattern (a quantum dot with two lithographically
defined gaps that act as tunnel barriers) and proves that patterning
of the hydrogenated SOI substrates with the STM tip and reasonable
resolution is possible.

\section{\label{sec:Alignment-and-STM}Alignment and STM based dopant device
fabrication on SOI substrates}

\begin{figure}
\begin{centering}
\includegraphics[width=16cm]{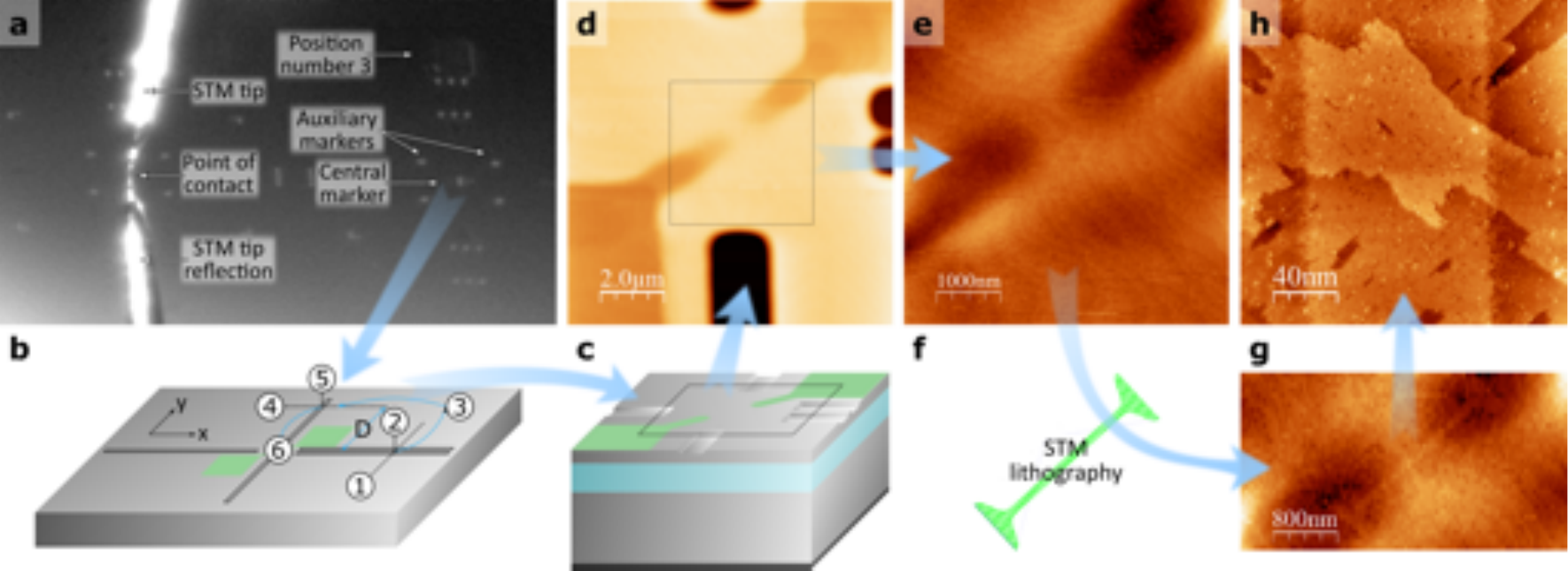}
\par\end{centering}
\caption{\label{fig:STM-image}Optical microscope image of the SOI sample with
alignment markers and STM tip visible. \textbf{a,} STM image of the
typical SOI surface. \textbf{b,c,} The procedure of localization of
the preimplanted contacts. STM images of device area before \textbf{d,e,
}and after \textbf{g,h,} the STM lithography step. }
\end{figure}

A key step in dopant device fabrication is in-situ alignment to pre-fabricated
structures, i.e. localization of the pre-implanted contacts with the
STM. Figure~\ref{fig:STM-image} shows the basic steps in the localization
and writing procedure.

First, the STM tip is brought close to the center of the desired device
position using an optical telescope attached to the STM system (Fig.~\ref{fig:STM-image}\textbf{a}).
The tip is positioned to the right of the device center and a full
range \SI{10}{\micro\m} line scan is made in the $y$-direction in
order to localize the alignment marker (1). Usually, the marker can
be found in the first scan, otherwise we adapt the coarse $y$ position
of the tip and make another scan. After centering the $y$-position
of the scan range to marker (2), we move the scan range by a defined
number of coarse steps (D) in the $y$ direction (3) and repeat the
procedure in order to find the alignment marker in the other direction
by long $x$-direction line scans (4). After centering the $x$-position
of the scanner range to the alignment marker, we move the tip backwards
in the $y$-direction ending up in the center of the device position
(6). Due to hysteresis in the coarse movement of the tip and piezo
creep effects there is typically an offset on the order of $1-2$\si{\micro\meter}
between the piezo scan range center and the real device position center.
With a \SI{10}{\micro\m} scanner range this is not a problem and
we typically have enough freedom to access the central area for device
patterning. In the future, the situation could be further improved
by using a closed-loop piezo scanner. 

A large overview scan is then performed (see Fig.~\ref{fig:STM-image}\textbf{d}).
This scan shows both the alignment markers (black areas) as well as
the finger implants (brown areas). These are easily recognizable thanks
to a depression that is a few monolayers deep and is a result of the
cleaning procedure after the ion implantation process (because of
slightly different etch rates of bare and implanted silicon). After
correct positioning of the STM scanning field (Fig.~\ref{fig:STM-image}\textbf{e}),
desorption of the desired pattern is performed (for details see Methods).
Figure~\ref{fig:STM-image}\textbf{g} shows an STM image of the device
position after desorbing a 120\,nm wide nanowire and Fig.~\ref{fig:STM-image}\textbf{e}
shows a detailed image of the desorbed pattern. The total duration
of the localization procedure is typically less than half an hour
(from introduction of the sample into the STM stage until the start
of the STM patterning) and the nanowire writing time, in this case,
was only about 20 minutes. After the STM patterning step, the standard
processing of the sample is performed, including dosing with phosphine,
activation anneal and overgrowth with 20\,nm of Si. 

\section{\label{sec:Weak-localization-in}Weak localization in dopant nanowires}

The quantum correction to the Drude conductivity in the 1D regime
is given by 

\begin{equation}
\delta G_{\mathrm{loc}}^{1\mathrm{D}}(B)=-\frac{e^{2}}{2\pi\hbar}\frac{1}{L}\left(\frac{1}{D\tau_{\phi}}+\frac{1}{D\tau_{B}}\right)^{-\frac{1}{2}}\label{eq:WL_1D_field_dep}
\end{equation}
where $e$ is electron charge, $\hbar$ is the reduced Planck constant,
$L$ is the length of the wire, $D$ is the diffusion constant, $\tau_{\phi}$
is the phase coherence time and $\tau_{B}$ is the magnetic relaxation
time that, for a given 1D geometry, can be calculated as 
\[
\tau_{B}=\frac{\hbar^{2}}{e^{2}}\frac{3}{W^{2}DB^{2}}
\]
where $W$ is the wire width (in the direction perpendicular to the
field and current) and $B$ is the magnitude of the magnetic field
perpendicular to the dopant device plane \cite{Beenakker1991}. Using
the definition of the phase coherence length $l_{\phi}=\sqrt{D\tau_{\phi}}$
we can rewrite equation \ref{eq:WL_1D_field_dep} as 
\begin{equation}
\delta G_{\mathrm{loc}}^{1\mathrm{D}}(B)=-\alpha\frac{e^{2}}{2\pi\hbar}\frac{1}{L}\left(\frac{1}{l_{\phi}^{2}}+\frac{e^{2}W^{2}B^{2}}{3\hbar^{2}}\right)^{-\frac{1}{2}}\label{eq:WL_1D_final}
\end{equation}
where $\alpha$ is a dimensionless correction factor that reflects
the inaccuracies in the geometry. The 1D assumption in the derivation
of equation \ref{eq:WL_1D_final} can be expressed by a condition
that $l_{\phi}\gg W$. If, on the other hand, $l_{\phi}\ll W$ we
need to apply a 2D weak localization theory which leads to a slightly
different expression for the correction of the conductance \cite{Beenakker1991}.

Equation \ref{eq:WL_1D_final} can be used to fit the conductance
of the dopant nanowires as a function of the magnetic field. Knowing
the geometry of the nanowires, the only free parameters are the phase
coherence length $l_{\phi}$ and the Drude conductivity $\sigma_{0}$,
which is the conductivity of the $\delta$-layer without the weak
localization correction, i.e. the limit of conductivity at high magnetic
fields. We also fit the prefactor $\alpha$ but since we assume that
it only reflects the inaccuracies in the wire geometry we restrict
the fit to a common value for all the temperatures. 

\begin{figure}[t]
\begin{centering}
\includegraphics[width=16cm]{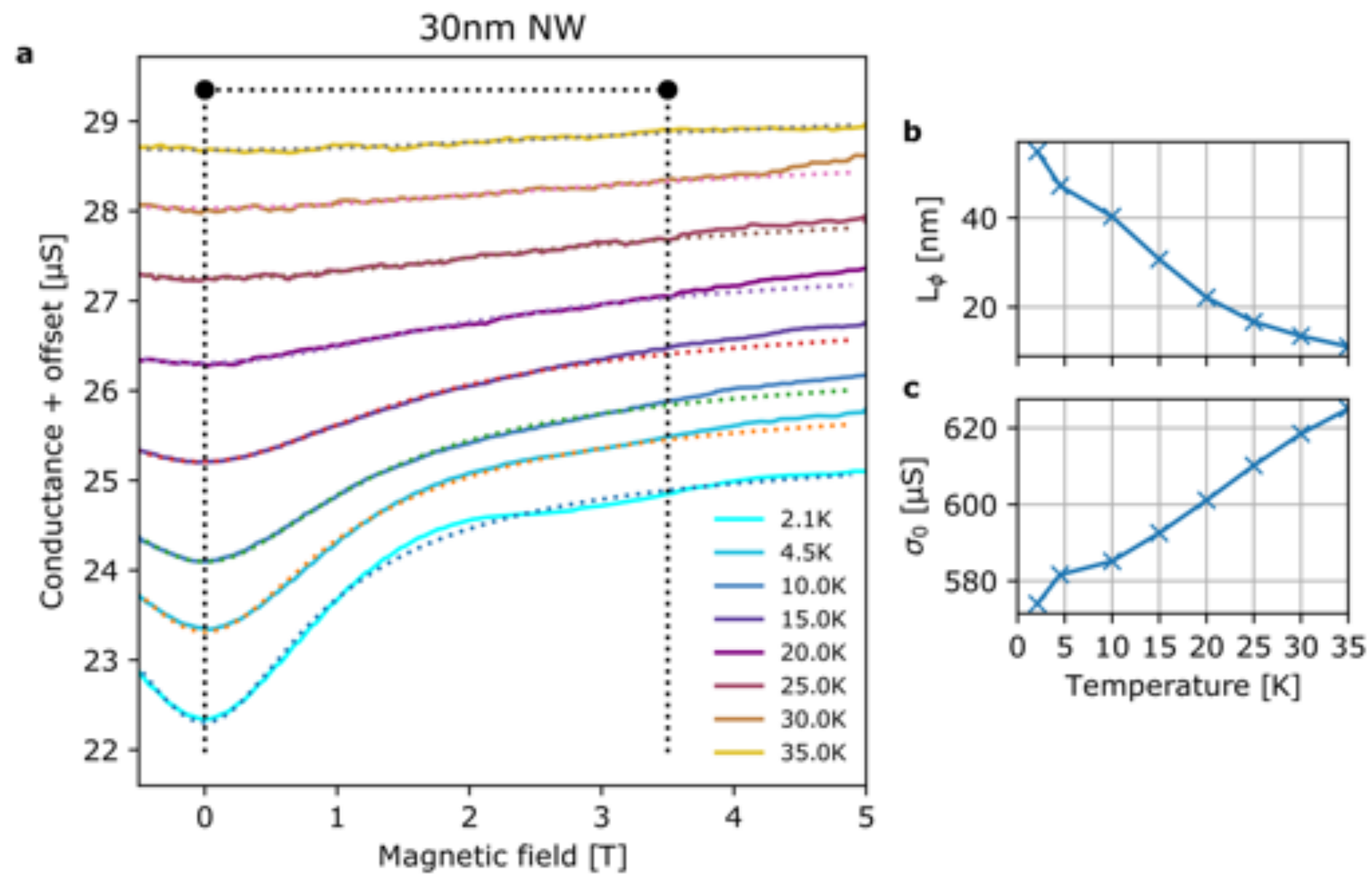}
\par\end{centering}
\caption{\label{fig:Magnetoresistance-30nmNW}\textbf{a}, Magnetoresistance
of the \textit{30\,nm\,NW} at different temperatures along with
theoretical curves obtained by fitting the 1D weak localization formula
\ref{eq:WL_1D_final}. Dotted black lines show the magnetic field
range used for fitting, since for large magnetic fields (above 3.5\,T)
pronounced universal conductance fluctuations affect the magnetoresistance.
The fit parameters $l_{\phi}$ and $\sigma_{0}$ for each temperature
are plotted in panels \textbf{b, c}.}
\end{figure}

\begin{figure}[t]
\begin{centering}
\includegraphics[width=16cm]{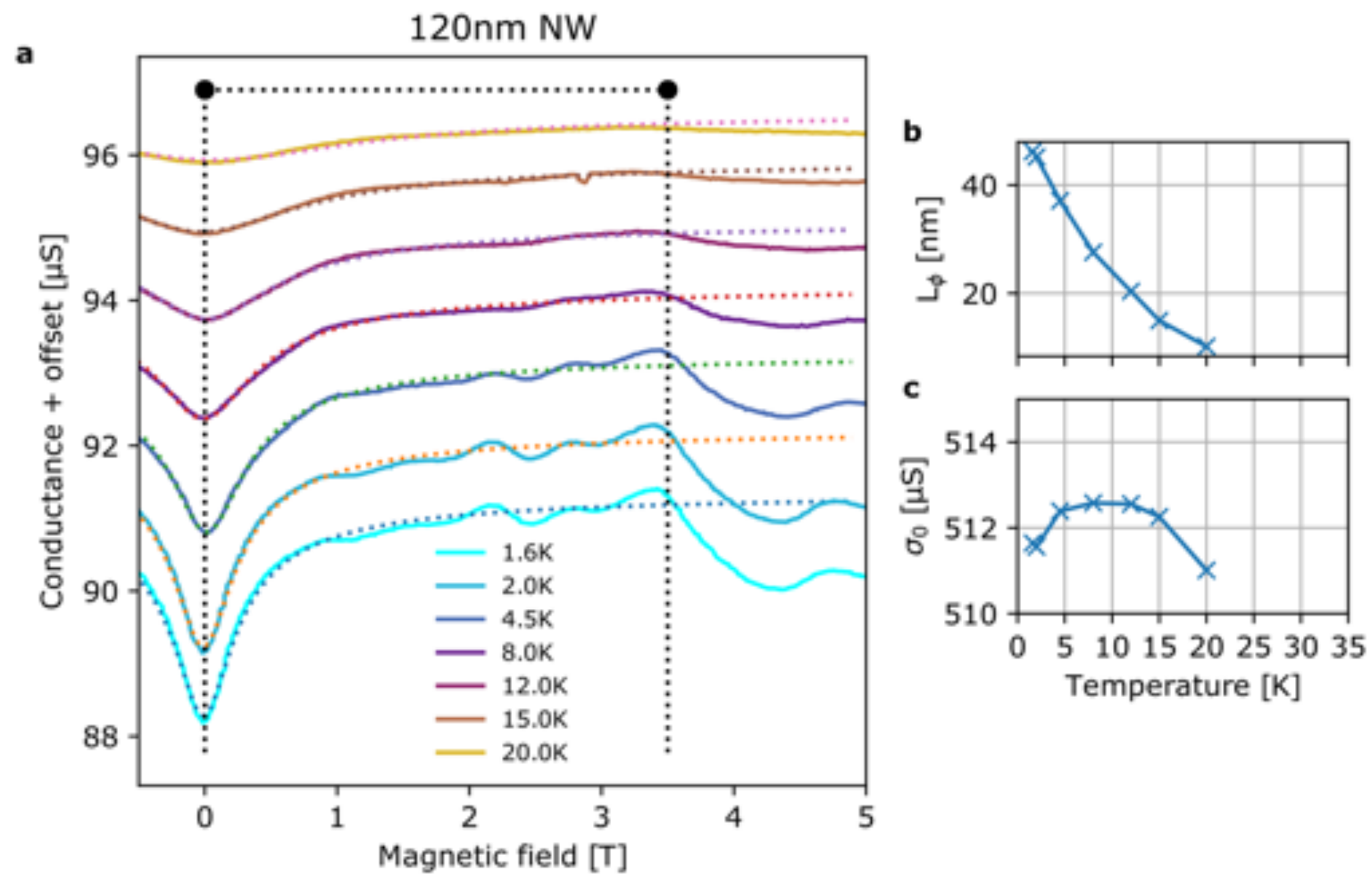}
\par\end{centering}
\caption{\label{fig:Magnetoresistance-120nmNW}\textbf{a}, Magnetoresistance
of the \textit{120\,nm\,NW} at different temperatures along with
theoretical curves obtained by fitting the 1D weak localization formula
\ref{eq:WL_1D_final}. Dotted black lines show the magnetic field
range used for fitting, since for large magnetic fields (above 3.5\,T)
pronounced universal conductance fluctuations affect the magnetoresistance.
The fit parameters $l_{\phi}$ and $\sigma_{0}$ for each temperature
are plotted in panels \textbf{b, c}. }
\end{figure}

Figures \ref{fig:Magnetoresistance-30nmNW} and \ref{fig:Magnetoresistance-120nmNW}
show the results of the fitting procedure for the two nanowires. The
value of the prefactor $\alpha$ is in both cases close to unity,
as expected (1.02 and 1.16 for the\textit{ 30\,nm NW} and \textit{120\,nm
NW}, respectively). In general, the shape of the experimental magnetoresistance
curves seems to be well captured by equation \ref{eq:WL_1D_final}.
The background conductivity $\sigma_{0}$ varies only slightly in
the temperature range and the phase coherence length decreases with
temperature in a similar fashion for both the nanowires, with a maximum
value of about 55\,nm at a temperature of 2\,K (similar to previously
reported value \cite{Ruess2007}). 

If we compare the phase coherence length $l_{\phi}$ to the width
of the nanowires $W$, the two values are essentially comparable for
both nanowires which means that the situation can neither be considered
1D nor 2D. In this intermediate regime an analytical formula for the
conductance correction is not easily obtainable. When we apply the
2D theory to the magnetoresistance curves, the quality of the fits
is substantially worse but the phase coherence length we obtain is
essentially the same. 

\begin{figure}[t]
\begin{centering}
\includegraphics[width=10cm]{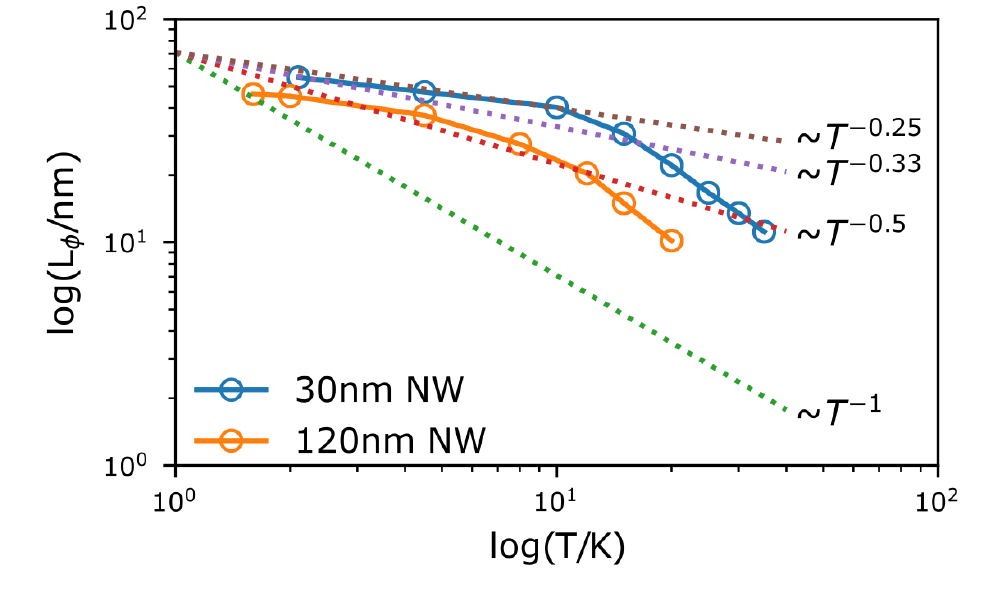}
\par\end{centering}
\caption{\label{fig:Magnetoresistance-lphi_comparison}Log-log plot of the
phase coherence length $l_{\phi}$ of the \textit{30\,nm} and \textit{120\,nm
NW}s as a function of temperature obtained from the fits to formula
\ref{eq:WL_1D_final}. Dotted lines show the slope of power law dependencies
with different exponents, for comparison. }
\end{figure}

In Fig. \ref{fig:Magnetoresistance-lphi_comparison} we compare the
phase coherence length for both nanowires. Its magnitude decreases
with increasing temperature but the dependence does not follow a simple
power law. At low temperatures for a 1D system, Nyquist dephasing
with temperature dependence $l_{\phi}\sim T^{-1/3}$ is believed to
be the dominant dephasing mechanism, while for higher temperatures
a crossover to a 2D behavior with a steeper decrease of the $l_{\phi}$
is expected \cite{ALTSHULER1985,Ruess2007}. This is qualitatively
consistent with the trend observed in our nanowires. For low temperatures
we also observe a saturation of the phase coherence length, similar
to \cite{Ruess2007}. 


\clearpage
